\documentclass[12pt,]{iopart}

\usepackage{iopams} 
\usepackage[breaklinks=true,colorlinks=true,linkcolor=blue,urlcolor=blue,citecolor=blue]{hyperref}
\usepackage{rotating}
\usepackage{graphicx}
\usepackage{epsfig}
\usepackage{epstopdf}

\begin{document}

\pagenumbering{arabic}

\title{Vortex dynamics and second magnetization peak in the iron-pnictide superconductor Ca$_{0.82}$La$_{0.18}$Fe$_{0.96}$Ni$_{0.04}$As$_2$} 
\author{I F Llovo$^1$, D S\'o\~nora$^1$, J Mosqueira$^1$, S Salem-Sugui Jr$^{2, \dagger}$, Shyam Sundar$^{2, \S}$, A D Alvarenga$^3$, T Xie$^4$, C Liu$^4$, S -L Li$^{4,5}$ and H -Q Luo$^{4,5}$}
\address{$^1$QMatterPhotonics Research Group, Departamento de Física de Partículas, Universidade de Santiago de
Compostela, Santiago de Compostela E-15782, Spain}
\address{$^2$Instituto de Fisica, Universidade Federal do Rio de Janeiro, 21941-972 Rio de Janeiro, RJ, Brazil} 
\address{$^3$Instituto Nacional de Metrologia Qualidade e Tecnologia, 25250-020 Duque de Caxias, RJ, Brazil}
\address{$^4$Beijing National Laboratory for Condensed Matter Physics, Institute of Physics, Chinese Academy of Sciences, Beijing 100190, P. R. China}
\address{$^5$Songshan Lake Materials Laboratory, Dongguan, Guangdong 523808, P. R. China}
\vspace{10pt}
\ead{$^\dagger$said@if.ufrj.br}
\ead{$^\S$shyam.phy@gmail.com}

\vspace{10pt}
\begin{indented}
\item[] June 2021
\end{indented}

\begin{abstract}
We report the studies of detailed magnetic relaxation and isothermal magnetization measurements in the vortex state of the 112-type iron-pnictide Ca$_{0.82}$La$_{0.18}$Fe$_{0.96}$Ni$_{0.04}$As$_2$ superconductor with $T_c$ $\sim$ 22 K. In the isothermal $M(H)$, a well defined second magnetization peak (SMP) feature is observed in the entire temperature range below $T_c$ for measurements with $H$ $\parallel$ $c$-axis. However, for $H$ $\parallel$ $ab$-planes, the SMP feature is suppressed at low temperatures, which might be due to 2D Josephson vortices forming at low temperatures and high magnetic fields in such an anisotropic system. A rigorous analysis considering the magnetic relaxation data for $H$ $\parallel$ $c$-axis suggests an elastic to plastic pinning crossover across $H_p$, which also seems accompanied with a possible phase transition in vortex lattice near $H_p$. Moreover, point disorder and surface defects are likely to be the dominant sources of pinning, which contribute to the $\delta l$-type of pinning in the sample. A high $J_c$, in access of 10$^5$ A/cm$^2$ observed could potentially make this material technologically important.
\end{abstract}
\vspace{2pc}

\maketitle

\section{Introduction}

Vortex physics in iron-pnictide superconductors is of a great importance for many future technological advancements \cite{Hosono:2018} as well as for the understanding of various exciting phases of vortex matter \cite{Kwok:2016}. Among the various interesting phenomenon observed in the vortex state of type-II superconductors, the second magnetization peak (SMP) phenomenon in the isothermal magnetization curves is one of them. Such phenomenon is ubiquitous in various low-$T_c$ conventional \cite{Lortz:2007, Stamopoulos:2004} as well as in high-$T_c$ unconventional superconductors \cite{Rosenstein:2005, Said:2020} and even in superconductors exhibiting multiple superconducting gaps, such as MgB$_2$ \cite{Stamopoulos:2004}. Also, the SMP has been investigated in nearly all families of iron-pnictide superconductors for magnetic field directions parallel and perpendicular to the crystal ab-plane \cite{Cheng:2019} (and references therein), its occurrence has not been observed in some pnictide crystals even with $H$ $\parallel$ $c$-axis, as for instance in overdoped Ba-KFe$_2$As$_2$ \cite{Song:2016}, in (Li-Fe)OHFeSe \cite{Wang:2017}, in La-doped CaFe$_2$As$_2$ \cite{Jung:2017} and in La$_{0.34}$Na$_{0.66}$Fe$_2$As$_2$ \cite{shyam:2019}. The importance of the SMP appearing on isothermal magnetization curves relies on its direct association with the peak effect appearing in the magnetic field dependence of critical current density, $J_c$($H$) \cite{Rosenstein:2010}, which is of technological importance. 

Despite it being observed in many superconductors, the mechanism responsible for the SMP is not totally understood and it appears to be material dependent \cite{Wang:2021}. For this reason the SMP has been studied in many different systems, usually by means of the vortex dynamics. Since the discovery of iron-pnictide superconductors \cite{Kamihara:2008} the SMP has been observed in many of these systems and explained in terms of a pinning-crossover as observed for Ba(Fe-Co)$_2$As$_2$, (Ba-K)Fe$_2$As$_2$, Ba(Fe-Ni)$_2$As$_2$, Ca(Fe-Co)As and (Ca-La)(Fe-Co)As$_2$ \cite{Shen:2010, shyam:2017a,Shyam:2017b, Shyam:2019a, Said:2010,Dawood:2017,Zhou:2016, Galluzzi:2019}, in terms of a phase transition in the vortex lattice as observed for Ba(Fe-Co)$_2$As$_2$,LiFeAs, and BaFe$_2$(As-P)$_2$ \cite{Kopeliansky:2010,Pramanik:2011, Said:2015, Miu:2020}, and also in terms of an order-disorder transition in some compounds \cite{Zehetmayer:2015, Miu:2012, Ionescu:2018}. It is to note that all the systems mentioned above share a similar layered crystal structure which consists of a spacer layer in-between of FeAs superconducting layers. However, the crystal structure in the 112 family has an additional spacer layer which leads to the enhanced spacing between FeAs superconducting layers \cite{Yakita:2014}. This contributes to the anisotropic superconducting properties in 112 family \cite{Zhou:2014, Xing:2016, Sonora:2017}. Since, the observed SMP in moderate anisotropic 1111-system is found to be due to the 3D order to 2D disorder phase transition \cite{Weyeneth:2009, Prozorov:2009}, it is interesting to find if a sample of 112-system with similar anisotropy as 1111-system, has the same origin of SMP. 

Here, we investigate the SMP and pinning behavior in a single crystal of the 112 type pnictide Ca$_{0.82}$La$_{0.18}$Fe$_{0.96}$Ni$_{0.04}$As$_2$ superconductor with superconducting temperature transition $T_c$ $\sim$ 22 K \cite{Xie:2017} and moderate anisotropy \cite{Sonora:2017}. A well pronounced SMP is observed in all isothermal $M(H)$ curves obtained with $H$ $\parallel$ $c$-axis even for temperatures very close to $T_c$ ($T$ = 20 K), and also for $H$ $\parallel$ $ab$-planes but only for temperatures above 14 K. As the SMP has been observed for $H$ $\parallel$ $c$-axis and $H$ $\parallel$ $ab$-planes in most of the pnictide superconductors \cite{Said:2011,Said:2013,Sharma:2013}, its absence for $H$ $\parallel$ $ab$-planes at lower temperatures (below 7.5 K) might be related to the anisotropic nature of the sample \cite{Sonora:2017} that might leads to the possible emergence of two dimensional Josephson vortices at low temperatures and high magnetic fields \cite{Said:2020, Moll:2012}. 

In order to study the possible origin of the SMP observed for H$\parallel$c-axis in our sample, extensive magnetic relaxation measurements were performed. For $H$ $\parallel$ $ab$-planes, magnetic relaxation data were within the noise level of the measurements, which prevented us from studying the vortex dynamics for this direction. The behaviour of the relaxation rate, $R$=dln$M$/dln(time), with field and temperature, as well as the dependence of the activation energy, $U_0$ = -$T/R$ \cite{Beasley:1969} with the critical current \cite{Feigel:1989} and of $U(M)$ with $M-M_{eq}$ \cite{Maley:1990, McHenry:1991, Abulafia:1996}, allowed us to study the vortex dynamics in the magnetic phase diagram of the system, and also to address the underlying mechanism for SMP. A crossover from collective (elastic) to plastic pinning has been observed across SMP, which is also accompanied with a possible phase transition in vortex lattice near $H_p$. Moreover, point disorder and surface defects seem to be the possible sources of vortex pinning in the sample. Self-field critical current density for $H$ $\parallel$ $c$-axis achieves $J_c$= 7$\times$10$^5$ A/cm$^2$ at $T$ = 2 K.  

\section{Experimental Details}

The Ca$_{0.82}$La$_{0.18}$Fe$_{0.96}$Ni$_{0.04}$As$_2$ single crystal used in this work was grown by a self-flux method as used in many other iron-based superconductors. A small crystal of mass, 0.306 mg with a roughly triangular platelet shape of surface area $S=0.65$ mm$^2$ and thickness $t=83.4$ $\mu$m (as determined from the density = 5.64 g/cm$^3$, calculated from the lattice parameters) was used in this study. A thorough description of the growth procedure can be seen in Ref.\cite{Xie:2017}. Details of the characterization by energy-dispersive x-ray spectroscopy (EDX) and x-ray diffraction can be seen in Ref.\cite{Sonora:2017} (crystal 11 of that reference). Let us just mention that it presents an excellent stoichiometric quality, with $x=0.176(3)$ and $y=0.045(3)$, and its diffraction pattern shows no spurious diffraction peaks, the $c$-axis lattice parameter being 10.348(1)~\r{A}.

The magnetization $M$ measurements were performed with a magnetic-properties measurement system (Quantum Design, model MPMS-XL) with magnetic fields up to 7 T applied both parallel and perpendicular to the crystal's $ab$ layers. For this purpose, a quartz tube was used as sample holder, to which the crystal was glued with GE varnish. In the case of $H\parallel c$, the crystal was glued to a $\sim$0.3 mm-wide slit, perpendicular to the quartz tube axis. Two plastic rods at the sample holder ends ensured an alignment of about $0.1^\circ$. $M(H)$ hysteresis curves were measured for both $H\parallel ab$ and $H\parallel c$, by using the MPMS's $hysteresis$ magnetic field charging mode. $M(t)$ relaxation curves were measured for $H\parallel c$ only (for $H\parallel ab$ the $M$ variation was of the order of the noise level). 

\section{Results and Discussion}

Figure 1 shows selected isothermal magnetization, $M(H)$, curves as obtained for $H$ $\parallel$ $c$-axis (a) and $H$ $\parallel$ $ab$- planes (b). All M$(H)$ curves shown in Fig. 1 are symmetric with respect to the magnetic field axis which reflects the dominant bulk pinning of the samples. The main information that can be extracted from these figures is that a pronounced second magnetization peak is observed for all $M(H)$ curves obtained with $H$ $\parallel$ $c$-axis. However, for $H$ $\parallel$ $ab$-planes the SMP is only observed in the temperature region above 7.5 K. As shown in Fig. 1b, the $M(H)$ curves below 14 K would show a SMP developing at increasingly higher fields at low temperatures, although below 7.5 K both branches of the $M(H)$ curves decreases monotonically. The disappearance of the SMP for $H$$\parallel$ $ab$-planes below 7.5 K is possibly related to the moderate anisotropy of the studied system, which may cause it to enter in a two dimensional regime for higher fields \cite{Sonora:2017}. We conjecture that the possible emergence of the Josephson vortices within the 2D regime would prevent the SMP to develop at low temperatures (below 7.5 K) due to the weaker pinning of the Josephson vortices than Abrikosov vortices \cite{Said:2020, Fehrenbacher:1992}. The curves of Fig. 1 (a) allow to extract the characteristic fields $H_{on}$, $H_p$ and $H_{irr}$ which are respectively the onset field of the SMP, the peak field, and the irreversible field. $H_{irr}$ was selected as the magnetic field at which the hysteresis amplitude becomes of the order of magnitude of the experimental noise. An example of $H_{irr}$ obtained by this criterion is shown in the inset of Fig. 1 (a). We observe that the SMP develops even for temperatures very close to $T_c$ for $H$ $\parallel$ $c$-axis direction. It is worth mentioning that the onset field $H_{on}$ and peak field, $H_p$ associated to the SMP for $H$ $\parallel$ $c$-axis are well defined in all isothermal $M(H)$ curves even at $T$ = 2 K.

\begin{figure}[h]
\centering
\includegraphics[height=10cm]{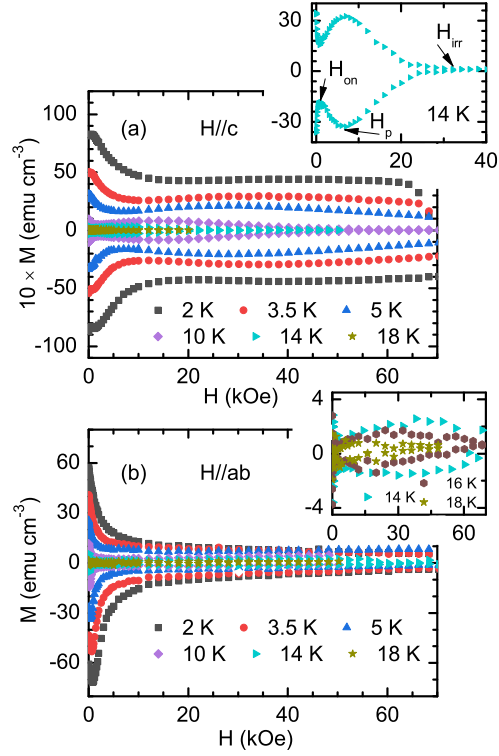}
\caption{\label{fig1} (a) Isothermal magnetic field dependence of the magnetization, $M (H)$, at selected temperatures well below $T_c$ for $H$ $\parallel$ $c$. In inset, the characteristic fields $H_{on}$, $H_p$ and $H_{irr}$ are well defined for $M(H)$ measured at $T$ = 14 K. (b) Isothermal $M (H)$ at selected temperatures for $H$ $\parallel$ $ab$. Inset shows the clear signature of SMP in isothermal $M (H)$ measured at 14 K, 16 K and 18 K.}
\end{figure}

In order to study the possible origin of the second magnetization peak in the sample, magnetic relaxation measurements were performed as a function of field for selected isothermal $M(H)$ curves, and as a function of temperature for selected applied magnetic fields. The study was only conducted for $H$ $\parallel$ $c$, as values of the magnetic moment during relaxation were   within the noise level observed for $H$ $\parallel$ $ab$- planes measurements. Magnetic relaxation curves were obtained for span times of about 80 minutes, and plots of ln$M$ vs ln(time) produced the usual linear curves from which the values of the relaxation rate $R$=dln$M$/dln(time) were extracted. Figure 2 shows plots of $M(H)$ curves obtained at $T$=10 K and 12 K, along with the respective relaxation rate curves obtained over each $M(H)$ curve from fields going from below $H_{on}$ up to above $H_p$. A change in the curvature of $R$ vs. $H$ can be observed for fields in the vicinity of $H_p$, the peak field, which might suggests a change in the pinning behaviour occurring near $H_p$. Moreover, a change in the curvature of each $R$ vs. $H$ near $H_{on}$ is associated to crossover of the single vortex pinning to collective vortex pinning \cite{shyam:2017a}. Similar peaks in the $R$ vs. $H$ curves, appearing between $H_{on}$ and $H_p$ have also been observed previously in other iron-pnictide superconductors \cite{shyam:2017a,Shyam:2019a} which have been attributed to a precursor phenomenon that leads to a SMP at higher fields \cite{Polichetti:2021}.

\begin{figure}[h]
\centering
\includegraphics[height=15cm]{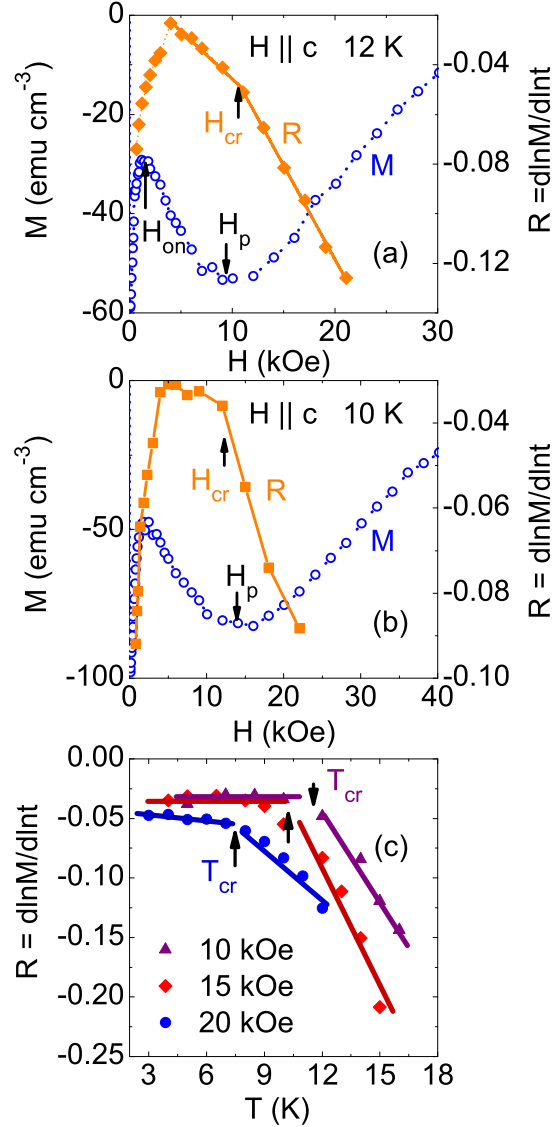}
\caption{\label{fig2} (a,b) Isothermal magnetic field dependence of the magnetization, $M(H)$, measured at $T$ = 12 K and 10 K, in the increasing magnetic field branch and the obtained relaxation rates, $R$, at fixed magnetic fields between $H_{on}$ and $H_p$ for respective $M(H)$. $H_{cr}$ indicates the change in slope in $R(H)$ which matches well with $H_{p}$. (c) Relaxation rate as a function of temperature,$R(T)$, for $H$ = 10 kOe, 15 kOe and 20 kOe. Arrows indicate, $T_{cr}$, the temperature where slop changes in each $R(T)$.}
\end{figure}

To check for the change in curvature observed in the curves of Fig. 2(a,b) for $H$ near $H_p$, magnetic relaxation measurements as a function of temperature for $H$=10, 15, and 20 kOe were performed. Figure 2(c) shows the resulting $R$ vs $T$ curves where a clear change in the behaviour of $R$ is observed near $T_{cr}$ (marked with an arrow). Later in the paper, it will be shown in  the $H$-$T$ phase diagram that the $T_{cr}$ obtained for each field is well matched with $H_p$ vs. $T$. 

As the $R$ vs. $T$, and $R$ vs. $H$ plots suggest a possible pinning crossover near $H_p$,  it would be interesting to see the behaviour of the activation pinning energy, $-T/R$ against 1/$J_{c}$, where $J_{c}$ is the critical current density. Figure 3 (a) shows selected curves of the magnetic field dependence of the critical current density, $J_c(H)$ for $H$$\parallel$$c$-axis as calculated using the Bean's critical-state model \cite{Bean:1962}. An expression for a triangular platelet with magnetic field parallel to $c$-axis was used to estimate the $J_c$ in the sample \cite{Poole}, $J_c(T, H) = \frac{15\Delta M(T, H)}{\sqrt{s^{-1}(s-a)(s-b)(s-c)}}$, where $a\approx1.25$~mm, $b\approx1.10$~mm, and $c\approx1.37$~mm are the sides of the triangle, $s=(a+b+c)/2$ is its semiperimeter, and $\Delta M(T, H)$ is the magnetization hysteresis. A peak, associated to the SMP in $M(H)$, is clearly visible for each $J_c(H)$ curve shown in Fig. 3 (a).

\begin{figure}[h]
\centering
\includegraphics[height=10cm]{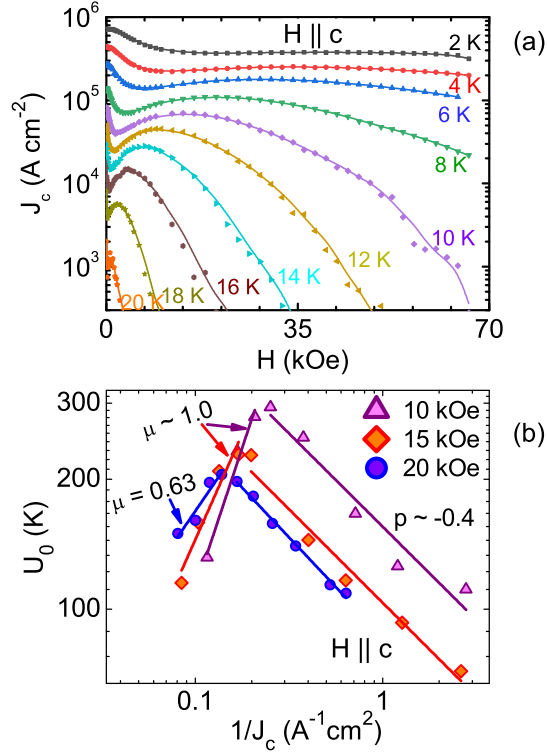}
\caption{\label{fig3} (a) Critical current density as a function of magnetic field, $J_{c}(H)$ obtained using Bean's critical state model \cite{Bean:1962} for $H$$\parallel$$c$. Solid lines are guide to eyes. (b) Activation energy, $U_0$ = T/R, as a function of inverse critical current density, 1/$J_c$, for $H$ = 10 kOe, 15 kOe, and 20 kOe. The exponents $\mu$ and $p$ found for each curve indicates an elastic pinning to plastic pinning crossover across SMP.}
\end{figure}

According to the collective pinning theory \cite{Feigel:1989} the activation energy $U$ $\sim$ (1/$J_c$)$^{\mu}$, where $\mu$ is a critical exponent. From this critical exponent, information about how the vortices are collectively pinned can be obtained. For instance, $\mu$=1/7 corresponds to single vortex, $\mu$=3/2 to small bundles of vortices and $\mu$=7/9 to large bundles. Figure 3 (b) shows two distinct behaviours occurring for all isofield curves where for lower values of the inverse critical current density, which corresponds to the region below $H_p$, the exponent $\mu$ = 1.07 and 0.98 for $H$ = 10 kOe and 15 kOe respectively and $\mu$ = 0.63 for $H$ = 20 kOe. Such values of $\mu$ are in agreement with a vortex lattice collectively pinned as small bundles and large bundles. The region corresponding to larger values of the inverse critical current density, which corresponds to the region above $H_p$, possesses a negative exponent which can not be explained in terms of the collective pinning theory. Such a region with a negative exponent \cite{Shyam:2017b, Zhou:2016} has been associated to plasticity of the vortex lattice, where the characteristic exponent is $p$=-0.5. As it can be observed in Fig. 3 (b) the exponent $p$ obtained for the three fields data is $p$$\sim$-0.4 ($p$=-0.4 for $H$=10,  $p$=-0.37 for 15 kOe and $p$=-0.46 for $H$=20 kOe) which agrees with the plastic exponent $p$=-0.5. Figure 4 suggests that the mechanism responsible for the second magnetization peak appearing in $M(H)$ curves in our sample is a crossover from collective to plastic pinning occurring as the SMP develops. It should be mentioned that such behavior, separating a low $J_c$ region from a higher J$_c$ region in $U_0$ vs. 1/$J_{c}$ isofield curves has also been observed previously in systems which do not exhibit the second magnetization peak \cite{shyam:2019}. To further study this change in behavior, a more rigorous analysis of the activation energy was performed, as first presented in Ref.\cite{Abulafia:1996}. 

The characteristic fields, $H_{on}$, $H_p$ and $H_{irr}$ obtained from the isothermal $M(H)$ curves for $H$ $\parallel$ $c$-axis are plotted as a function of temperature in the $H$-$T$ phase diagram, shown in Fig. 4. Contrary to the $H_{on}$ line, the temperature dependence of $H_p$ and $H_{irr}$ follow $\sim$ $(1-(T/T_c))^n$ behavior, where, $n$ = 2, 1.3 are obtained for $H_p$ and $H_{irr}$ lines respectively. A similar temperature dependence of the irreversibility line was also observed in Ref. \cite{Yeshurun:1988}. It is interesting to note that at low T, the $H_{on}$ and $H_p$ follow relatively opposite curvatures with temperatures, which may lead to the merging of $H_p$ and $H_{on}$ lines at temperatures below 2 K. A merging of the $H_{on}$ line with the $H_p$ line at high fields would imply the disappearance of the SMP, which in the present case of a highly anisotropic system could be associated to a possible field-induced crossover to a two dimensional vortex system. It should be mentioned that the vortex physics associated to the $H_{on}$ and $H_p$ are different in nature, where, $H_{on}$ is reported to be associated to a crossover from single vortex-pinning to a collective vortex-pinning regime \cite{Shyam:2017b, Said:2010, Abulafia:1996}. This supports a change in pinning strength from weak to strong across $H_{on}$ as reported in ref. \cite{Galluzzi:2018}. On the other hand, different mechanisms have been reported to be responsible for the SMP at $H_p$ in different systems, as discussed in the introduction. The origin of the SMP in present sample is discussed in the later part of the paper. We also plot in the phase diagram the corresponding $H_{cr}$ and $T_{cr}$ values of the kinks observed in $R$ vs. $H$ and $R$ vs. $T$ plots respectively. It is interesting to note that the $H_{cr}$ and $T_{cr}$ points in the $H-T$ phase diagram lie almost perfectly on the $H_p$ line. This fact, despite evidencing a change in the pinning mechanism, as discussed above, may also be related to a possible vortex lattice phase transition taking place as the second magnetization peak develops \cite{Rosenstein:2005, Rosenstein:2007}, which deserves further investigation. Moreover, a contour plot of $J_c$ obtained at various temperatures below $T_c$ in the magnetic field range 0-7 T is also shown in Fig. 4, to track the $H_{on}$, $H_p$ and $H_{irr}$ with the changes in $J_c$($H, T$).

\begin{figure}[h]
\centering
\includegraphics[height=8cm]{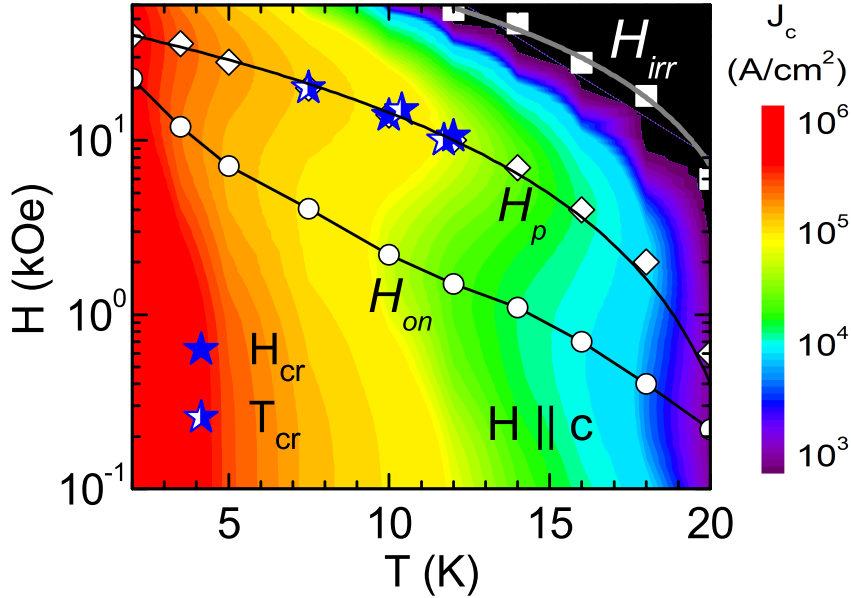}
\caption{\label{fig5} An $H$-$T$ phase diagram representing the characteristic fields, $H_{on}$, $H_{p}$ and $H_{irr}$. Solid lines in $H_{irr}$ and $H_{p}$  are fit to the data as explained in the text. $H_{cr}$ and $T_{cr}$ are field and temperature marked in Fig. 2. Solid line in $H_{on}$ is guide to the eyes. The critical current density as a function of magnetic field at different temperatures is plotted in a contour plot.}
\end{figure}

The activation energy, $U(M)$, was obtained following an approach developed by Maley et al. \cite{Maley:1990, McHenry:1991} assuming that the isofield $U(M)$ curves measured at different temperatures should be a smooth function of $M-M_{eq}$, where $M_{eq}$ is the equilibrium magnetization. It should be mentioned that for our system $M-M_{eq}$ $\approx$ $M$. In this approach, $U(M)$=-$T$ln($dM/dt$)+$C$$T$, where $C$ is an intrinsic constant. $U(M)$ is then calculated from the isofield relaxation curves obtained for several temperatures and plotted as a function of $M$. The appropriate constant $C$ for the system defines the smooth curve. But as pointed out in Ref.\cite{McHenry:1991}, for most systems the smooth curve is only obtained by dividing $U(M)$ by a scaling function $g$($T$/$T_c$) which carries the behavior of the coherence length $\xi$$(T)$. Figure 5 shows a plot of $U$/(1-$T$/$T_c$)$^{3/2}$ vs. $M$ for $H$ = 10 kOe exhibiting a smooth behavior with $M$, which was obtained for $C$ $=$ 15. Similar curves with the same constant $C$ were obtained for $H$ $=$ 15 and 20 kOe (not shown here for brevity). With the value of $C$ = 15 we can calculate $U(M)$ from the isofield magnetic relaxations obtained on the isothermic $M(H)$ curves for fields below and above the $H_p$. Figure 6 (a) shows the $U(M)$ curves obtained for isothermal $M(H)$ at 12 K where a clear change in the behavior of $U(M)$ is observed as $H_p$ is crossed. To check for a possible pinning crossover across $H_p$, we exploited an expression from the collective pinning theory, $U(B,J)$ = $B^{\nu}$$J^{-\epsilon}$ $\approx$ $H^{\nu}$$M^{-\epsilon}$, where, $\nu$ and $\epsilon$ are exponents for specific pinning. Figure 6 (b) shows plots of selected $U(M)$ curves of Fig. 6 (a) after being scaled by $H$$^{\nu}$, as in Ref. \cite{Abulafia:1996}. For $H$ below and above $H_p$, the scaling was obtained for $\nu$ = 0.5 and -0.5 respectively. A positive value of exponent $\nu$ supports the collective pinning for $H$ $<$ $H_p$ \cite{Feigel:1989, Abulafia:1996}, whereas a negative $\nu$ exponent for fields above $H_p$ supports plastic pinning. Although, the expected values of $\nu$ are 1 and -0.5 for collective and plastic pinning respectively, exponent $\nu$ smaller than 1, associated to collective pinning, have also been observed in other systems \cite{Said:2020, Shyam:2017b, Shyam:2019a}. The inset of Fig. 6 (b) shows the corresponding $M(H)$ at 12 K evidencing $H_p$. The plots of Fig. 6 (b) clearly demonstrate that the mechanism explaining the second magnetization peak in our sample is a crossover from collective to plastic pinning occurring as the peak field $H_p$ is crossed. As discussed in the last paragraph, there is also a possibility of vortex lattice phase transition below $H_p$. Such change in vortex lattice near $H_p$ may creates an energetically favorable scenario for the plastic pinning at fields above $H_p$. Similar behavior has also been observed in the case of Co-doped 122 iron pnictide superconductor, where a a vortex lattice phase transition below $H_p$ accompany the collective to plastic creep crossover across $H_p$ \cite{shyam:2017a, Kopeliansky:2010, Prozorov:2008}. Moreover, Kopeliansky $et$ $al$ \cite{Kopeliansky:2010} also stated that such a crossover (collective-plastic) in vortex dynamics may accompany a thermodynamic phase transition in vortex lattice, as seen previously in case of YBCO \cite{Abulafia:1996, Deligiannis:1997}. However, any direct observation of vortex lattice phase transition below $H_p$ and its correlation with SMP is yet to be confirmed in iron pnictide superconductors. 

\begin{figure}[h]
\centering
\includegraphics[height=8cm]{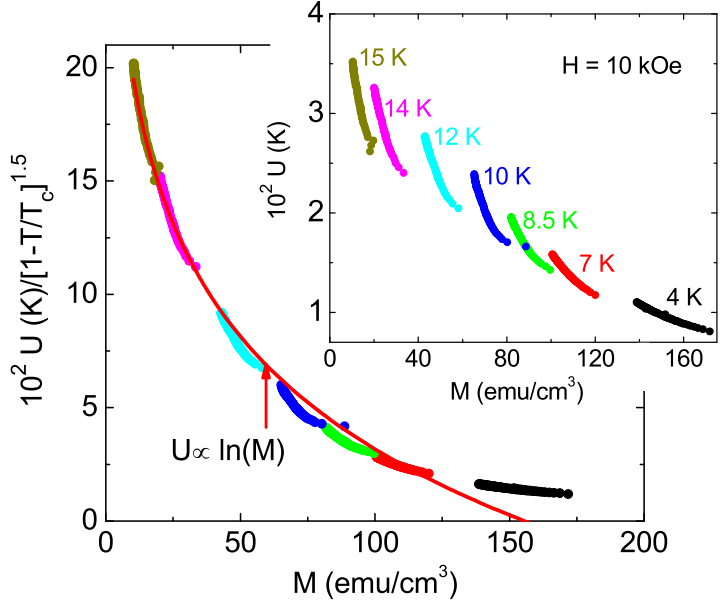}
\caption{\label{fig6} $U/g(T/T_c)$ vs. $M$ scaled plot obtained for $H$ = 10 kOe using Maley's criterion explained in the text. Solid line shows the ln($M$) dependence of scaled curve. Inset shows the $U(M)$ without scaling using $g(T/Tc)$ function. }
\end{figure}

\begin{figure}[h]
\centering
\includegraphics[height=13cm]{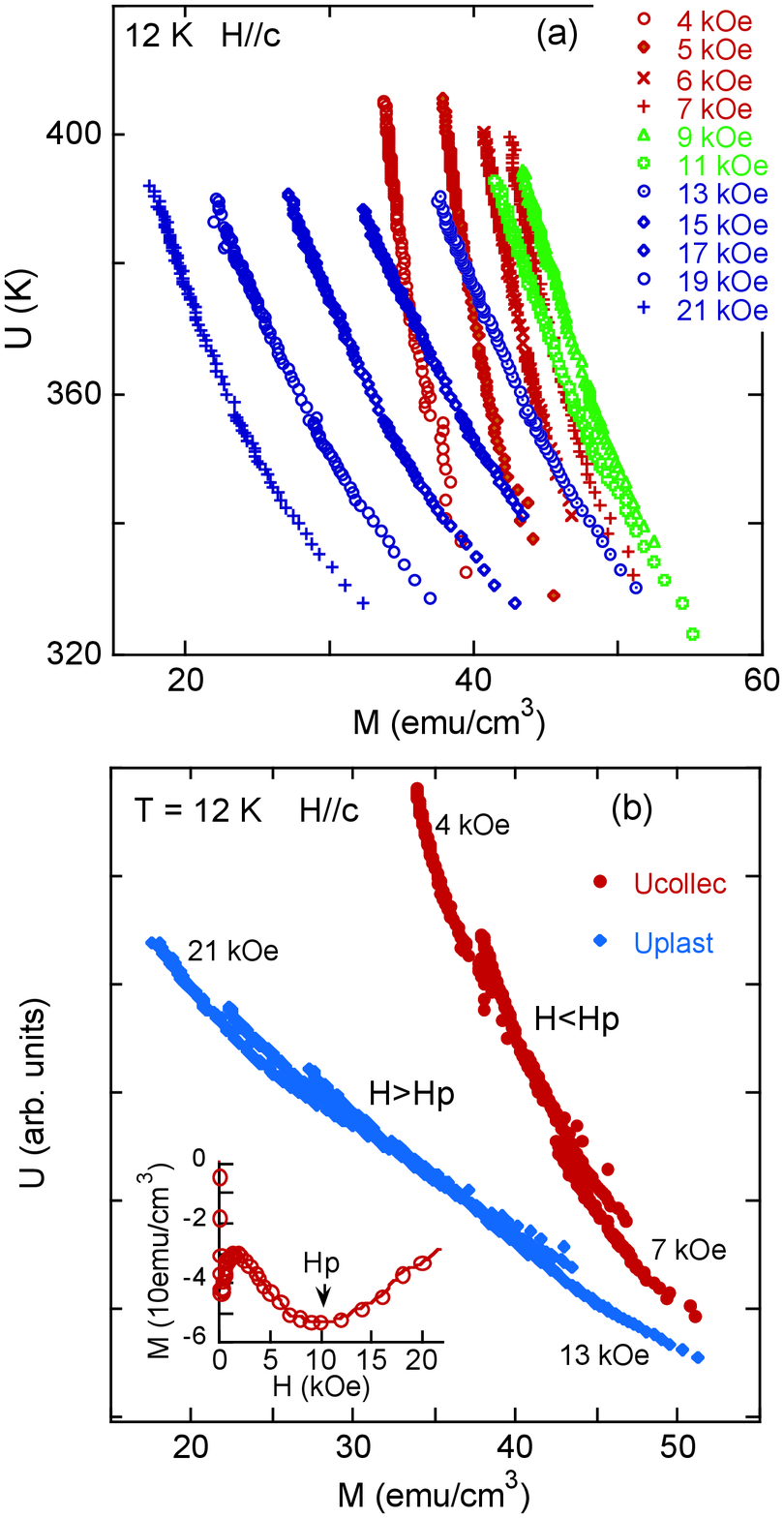}
\caption{\label{fig7} (a) Activation energy, $U$, vs. magnetization, $M$, for different magnetic fields above and below the $H_p$ at $T$ = 12 K. (b) Scaled $U$ curves using collective pinning theory show elastic pining for $H$ $<$ $H_p$ and plastic pinning for $H$ $>$ $H_p$. Inset shows the initial branch of $M(H)$ measured at $T$ = 12 K for $H$ $\parallel$ $c$-axis.}
\end{figure}

Figure 7(a) shows a plot of the so called remanent critical current density normalized by its value extrapolated to $T$=0, $J_c$($T$)/$J_c$(0) against temperature for $H$ $\parallel$ $c$-axis. The remanent critical current corresponds to the critical current at zero field extracted from the Bean's model, where $\Delta M$ is obtained by subtracting the magnetization for $H$=0 belonging to the decreasing field branch from the magnetization curve for $H$=0 belonging to the increasing field branch after cycling the correspondent isotherm $M(H)$ in negative magnetic fields. For comparison, Fig. 7(a) also shows the values obtained from a model developed in Ref.\cite{Griessen:1994} which considers two possible pinning of the type $\delta l$, for which $J_c$($T$)/$J_c$(0) $\sim$ (1-$t^2$)$^{5/2}$(1+$t^2$)$^{-1/2}$, and $\delta T_c$, for which  $J_c$($T$)/$J_c$(0) $\sim$ (1-$t^2$)$^{7/6}$(1+$t^2$)$^{5/6}$, where $t$ = $T$/$T_c$. As observed in other pnictides \cite{Shyam:2017b} and references therein, the dominant pinning in our sample follows neither of the above two behaviors considered in Ref.\cite{Griessen:1994}. However, such behavior of $J_c(T)$ may be explained considering the weak and strong pinning at low and high magnetic fields respectively, as shown in case of FeSe$_{0.5}$Te$_{0.5}$ superconductor \cite{Galluzzi:2019b}.

\begin{figure}[h]
\centering
\includegraphics[height=12cm]{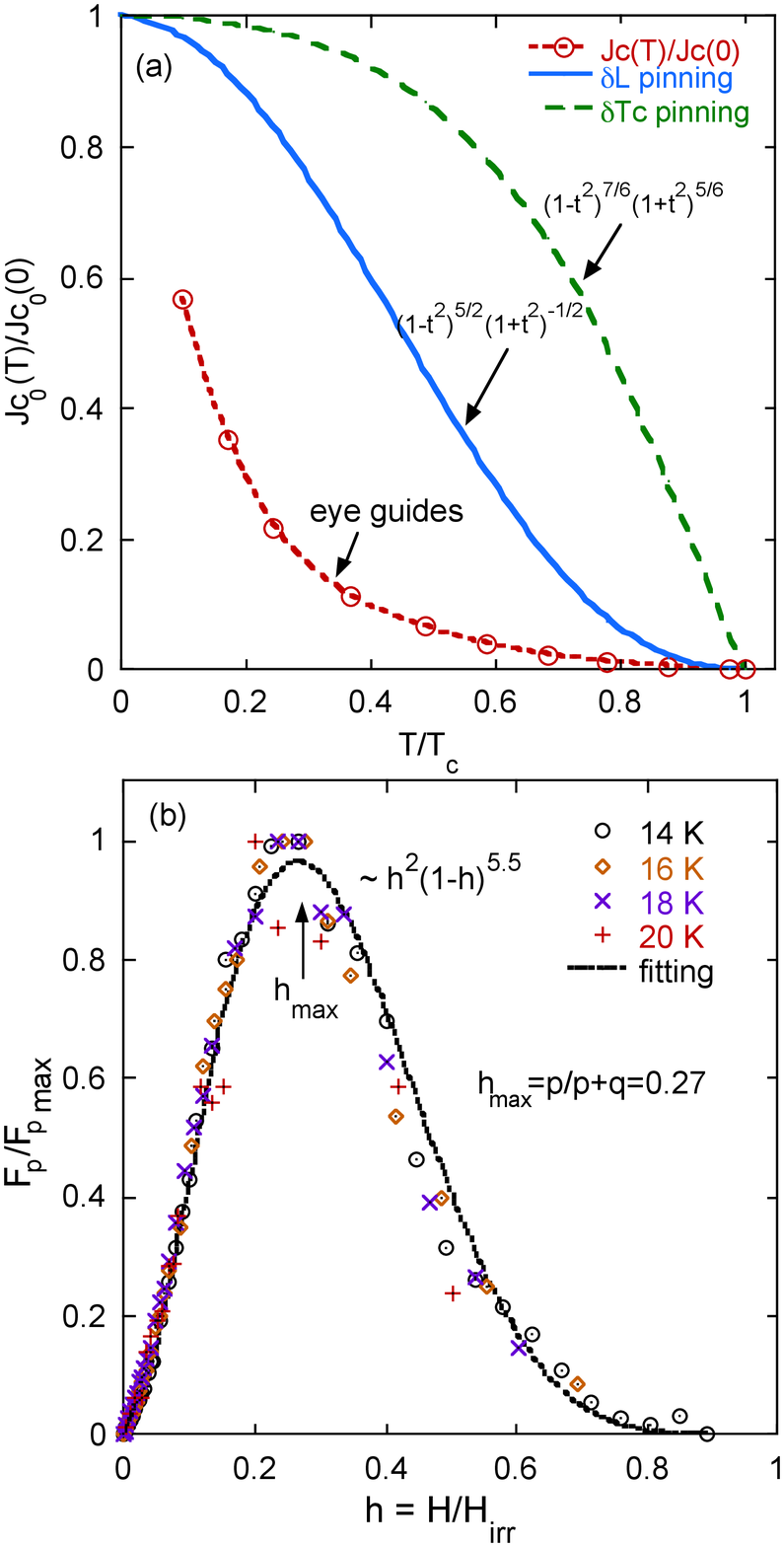}
\caption{\label{fig8} (a) Normalized critical current density, $J_c$($T$)/$J_c$(0) as a function of reduced temperature, $t$ = $T/T_c$ from experimental data. Solid lines represents the $\delta l$ and $\delta T_c$ pinning using models in Ref. \cite{Griessen:1994}. (b) Normalized pinning force density, $F_p$($H$)/$F_p$$_{max}$, as a function of the reduced field, $h$ = $H/H_{irr}$ for different temperatures. The solid line is the fitting of scaled curves using the expression, $F_p$($H$)/$F_p$$_{max}$ $\sim$ $h^p$(1-$h$)$^q$, where, $p$ and $q$ are fitting parameters \cite{Dew-Hughes:1974}. Details of the fit are given in the main text.}
\end{figure}

Figure 7(b) shows a plot of the normalized pinning force density, $F_p$($H$)/$F_p$$_{max}$, as a function of the reduced field $h$=$H$/$H_{irr}$, for several temperatures. The pinning force density is obtained using expression, $F_p$=$H$$\times$$J_c$. The scaling of the different isothermal $F_p$($H$)/$F_p$$_{max}$ curves with $h$=$H$/$H_{irr}$ is a powerful tool \cite{Dew-Hughes:1974, Koblischka:2016} commonly used in new materials to identify the dominant pinning acting within certain temperature regions \cite{Zhang:2017}. The importance of this plot also relies on the identification of the magnetic field region, $h_{max}$, where the pinning force presents its maximum, which might be important for application purposes \cite{Dew-Hughes:1974, Koblischka:2016}. As shown in Fig.7 (b), all curves collapse in one, with a very clear maximum appearing for $H$/$H_{irr}$=0.27. The solid curve in this figure is a best fit to the well known Dew-Hughes expression, $F_p$($H$)/$F_p$$_{max}$ $\sim$ $h^p$(1-$h$)$^q$, where values of $p$ and $q$ are associated to the characteristic types of pinning explained in the model \cite{Dew-Hughes:1974}. The best fit shown in Fig. 7(b) was obtained with $p$=2 and $q$ = 5.5 where $h_{max}$= $p$/($p$+$q$) $\sim$ 0.27, which is consistent with the $h_{max}$ observed from data. It is important to mention that in Dew-Hughes model \cite{Dew-Hughes:1974, Koblischka:2016}, $\delta l$ pinning due to point disorder expects, $p$ = 1 and $q$ = 2, with $h_{max}$ = 0.33. Therefore, in our case, $h_{max}$ = 0.27 indicates the pinning due to point disorder. However, higher values of $p$ and $q$ and slightly lower $h_{max}$ than what is ideally expected for Dew-Hughes model, suggests additional pinning at play, likely due to the surface defects. Similar results have been observed in other iron pnictide superconductors as reported in Refs. \cite{Shyam:2019a, Shahbazi:2013, Shahbazi:2013a, Gennep:2020}. Moreover, according to Dew-Hughes model \cite{Dew-Hughes:1974}, $h_{max}$ $<$ 0.5 indicates a $\delta l$ pinning, while $h_{max}$ $>$ 0.5 suggests $\delta T_c$ pinning, therefore, $h_{max}$ = 0.27 in present case is suggestive of $\delta l$ pinning. In addition, more than one pinning centers at play may lead to the deviation from the $\delta l$ pinning model, as observed in Fig. 7(a).

\section{Conclusions}

In conclusion, we investigated the second magnetization peak (SMP) and the associated pinning properties in a single crystal of the iron-pnictide Ca$_{0.82}$La$_{0.18}$Fe$_{0.96}$Ni$_{0.04}$As$_2$ superconductor. In isothermal $M(H)$ measurements for $H$ $\parallel$ $c$-axis, the SMP was observed for the entire temperature range below $T_c$. However, the SMP was observed only for temperatures close to $T_c$ for $H$ $\parallel$ $ab$-planes. A detailed investigation based on Maley's analysis and collective pinning theory suggests that the SMP in the sample may be explained in terms of an elastic pinning to plastic pinning crossover across $H_p$, which also seems accompanied with a possible vortex lattice phase transition. However, any direct observation of such phase transition in vortex lattice near $H_p$ and its correlation with SMP is yet to be confirmed. For $H$ $\parallel$ $ab$-planes, the suppression of the SMP at low temperature may be related to the sample anisotropy, which in turn leads to the 2D Josephson vortices at low temperature and high magnetic fields. Based on the Dew-Hughes model, pinning analysis for $H$ $\parallel$ $c$ suggests that point disorder in addition with surface defects are the possible sources of vortex pinning, which are in favor of a $\delta l$-type pinning in the sample. Moreover, the critical current density has been found to be higher than 10$^5$ A/cm$^2$ for temperatures below 8 K in the entire magnetic field range of the measurements. This property makes this compound technologically relevant for use in high magnetic field generation.

\section*{Acknowledgements}
IFL, DS, and JM acknowledge support by the Spanish Agencia Estatal de Investigación (AEI) and Fondo Europeo de Desarrollo Regional (FEDER) through project PID2019-104296GB-I00, and by Xunta de Galicia (grant GRC no. ED431C 2018/11). IFL acknowledges financial support from Xunta de Galicia through grant ED481A-2020/149. Authors would like to thank the use of RIAIDT-USC analytical facilities. SSS and ADA acknowledges support from CNPq. This work at IOP, CAS is supported by the National Key Research and Development Program of China (Grants No. 2018YFA0704200, No. 2017YFA0303100, and No. 2017YFA0302900), the National Natural Science Foundation of China (Grants No. 11822411, No. 11961160699, and No. 11874401), the Strategic Priority Research Program (B) of the Chinese Academy of Sciences (CAS) (Grants No. XDB25000000 and No. XDB07020300) and K. C. Wong Education Foundation (GJTD-2020-01). HL is grateful for the support from the Youth Innovation Promotion Association of CAS (Grant No. Y202001).

\section*{References}
\providecommand{\newblock}{}

\end{document}